\documentclass[prl,amsmath,twocolumn,showpacs]{revtex4}

\usepackage{epsfig}

\begin{document}

\title{Residence time statistics for $N$ blinking quantum dots and other stochastic processes}
\author{S. Burov, E. Barkai}
\affiliation{Department of Physics, Institute of Nanotechnology and Advanced Materials,  Bar Ilan University, Ramat-Gan
52900, Israel}

%
%
\begin{abstract}
We present a study of residence time statistics for  
$N$ blinking quantum dots.
With numerical simulations and exact calculations  
we show sharp transitions for a critical
number of dots. 
In contrast to expectation the fluctuations in the
limit of $N \to \infty$
are  non-trivial.  
Besides quantum dots our work describes residence
time statistics in several other many particle systems for example
 $N$  Brownian particles.  
Our work provides a natural framework to detect non-ergodic kinetics from 
measurements of many blinking chromophores, without the need to reach the
single molecule limit.   
\end{abstract}


\maketitle

\begin{figure}
\begin{center}
\epsfig{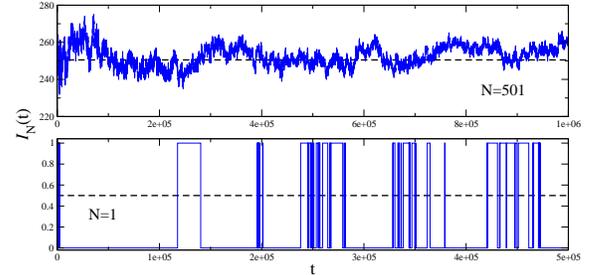}
\end{center}
\caption{ 
Model two state blinking QD dot with $\alpha=1/2$.
For $N=1$ (lower panel) power law
distributed sojourn times in bright $+1$ and dark $0$ states lead to 
long residence times, of the order of the measurement time, in both states. 
Upper panel exhibits a superposition of the signal from $N=501$ 
processes. In this Letter we investigate the fluctuations of the 
residence time, namely the time the signal remains above the
ensemble mean (the dashed line) 
 showing they are not trivial even in the limit of $N \to \infty$.
}
\label{fig1}
\end{figure}

 Residence time statistics has attracted considerable interest
as a tool describing non-equilibrium phenomena. For a fluctuating
time trace ${\cal I}(t)$ the residence time $t^{+}$ is the total time
the signal remains beyond some threshold. A classical result in this
field is L\'evy's arcsine law 
\cite{Godreche,Majum}. 
Then $t^{+}$ is the residence  time of a one dimensional 
Brownian particle in half space.
 In contrast with naive expectation the PDF  of $t^{+}/t$  
($t$ is the measurement time) is not centered on its mean, instead
it has a $\cup$ shape with singularities described
by the persistence exponent $\alpha=1/2$ for simple diffusion 
\cite{Majumdar} (see details below). 
Similar bi-modal distributions were
obtained for random walks in disordered environments
\cite{Refm},
the persistence problem of diffusion in dimension $d$ \cite{Newman},
a melting heteropolymer Sinai model \cite{Oshanin}
 and
for the fluorescence
signal from a quantum dot (QD) \cite{Dahan,GenadPRL}.
A {\em single} QD when interacting with 
a continuous wave laser field exhibits blinking with 
power law kinetics  which is related to non trivial
residence times, non-stationary and non ergodic behavior 
\cite{Dahan,GenadPRL,Marcus,PhysToday}. 
At random times the dot will switch from  a bright
state where many photons are emitted to a dark state (see Fig. \ref{fig1}). 
Probability density
function (PDF) of on and off times are power laws  $\psi(\tau) \sim 
\tau^{-(1 + \alpha)}$ at least for low laser intensity \cite{Kuno}. 
 In most cases
$\alpha\simeq 1/2$ though $1/2<\alpha<1$ was also reported. Most interestingly
the average on and off times $\langle \tau \rangle=\int_0 ^\infty \tau \psi(\tau) {\rm d} \tau$ diverges, i.e. the
dynamics is scale free. Several physical models explaining this behavior
were suggested \cite{Marcus,PhysToday}, for example based on 
normal diffusion \cite{GenadPRL}, however so
far no experimental smoking gun provides confirmation to a specific
mechanism responsible for  the power law
blinking. Further the effect is not
limited to QDs, it is found also  for organic
single molecules provided
that they are embedded in a disordered material \cite{PhysToday}. 
One can rightly wonder
why after many decades of spectroscopy we find surprises once individual
objects are detected \cite{Chung}. 
Can we not infer the strange  blinking from measurements
of many dots/single molecules? 
In particular can the broken ergodicity previously
measured for an individual
dot \cite{Dahan,GenadPRL} be detected also for an ensemble of dots? 

 For that aim we investigate the problem of residence time statistics
for $N$ blinking dots. We will soon show that this problem
 can be mapped onto
an interesting  many particle Brownian problem and other stochastic processes. 
Surprisingly non trivial residence time
statistics is found here even in the limit of
$N\to \infty$. This is  an important  indication that by the analysis of
light emitted from an ensemble of chromophores we may detect
the strange kinetics without the need to reach the single molecule
limit. 
We investigate transitions in the PDF
of residence times (from $\cup$ to $\cap$ shape, as explained below)
which are controlled by the number of dots $N$. 
We show that a critical number of dots
marks a 
border line between 
the single particle domain and the many particle limit. 
Surprisingly we find values of $\alpha$ below which 
$\cup$ shape PDFs are robust, namely for certain $\alpha$
the $\cup$ shape PDF is found even for $N\to \infty$.
Roughly speaking this is an important indication that elements
of L\'evy's arcsine law survive the $N\to \infty$ limit, hence persistence
exponents \cite{Majumdar}
from signals composed of a super-position of many
elements are more common than expected. 
Our main analytical calculation is an exact expression for the
variance of the residence time which is used to identify
transitions between sub and super uniform statistics defined below.


 {\em Model and observable.} 
We consider $N$ independent though statistically identical
QDs. Each dot undergoes the following simple renewal process.
The dot can be in a bright state where the intensity of light is $I_i(t)=1$
or a dark state with intensity $I_i(t)=0$. Sojourn times in state on and off
are independent identically distributed random variables with a common
PDF $\psi(\tau)$. $\psi(\tau)$ is moment less, namely for large $\tau$
$\psi(\tau) \sim A \tau^{-(1 +\alpha)}$ and $0<\alpha<1$. 
The signal of $N$ dots is
${\cal I}_N (t) = \sum_{i=1} ^N I_i(t) $ whose  
ensemble mean is obviously $\langle {\cal I}_N \rangle= N /2$. 
Here  we 
investigate the 
time $t_{+}$ the process remains above its mean (see Fig. \ref{fig1}). 
The residence fraction is defined as $0<p^{+}= t^{+} / t<1 $ where 
$t$ is the total measurement time. 

{\em Relation with Brownian motion.} The model
and its variants
describe many physical systems and processes beyond QDs
e.g. well known L\'evy walks \cite{Shles}
which describe tracer diffusion in turbulent flow, certain
chaotic systems \cite{Zumofen} and recently  models of
$1/f$ noise \cite{Grig1,Kantz}.
Probably the best well known example 
is Brownian motion. 
Consider a single Brownian particle in one dimension and
let $I_1(t)=1$
for a particle being in $x>0$ and $I_1(t)=0$ for $x<0$.
As well known the PDF of first passage times and hence of sojourn
times in $x<0$ or $x>0$ follow power law behavior $\psi(\tau) \propto \tau^{- 3/2}$
and hence $\alpha =1/2$. 
For a single particle the time  $t^{+}$
 is simply the residence time in
half space $x>0$  which follows the classical arcsine distribution 
of P. L\'evy 
(see details below).
For $N$  Brownian particles $t^{+}$ is
the time the majority of $N$ particles are on $x>0$. 
Naively one would expect that this time
in the $N\to \infty$ limit would be equal to half of the measurement time.
This turns out to be wrong and features of L\'evy's arcsine law survive
the $N \to \infty$ limit.

 {\em Known results for a single QD $N=1$.} 
Lamperti obtained the PDF of the residence fraction  $0<p^{+}<1$ 
\cite{Godreche,GenadPRL} 
\begin{equation}
 l_a(p^{+}) = 
{ \left[\sin\left(\pi \alpha\right) /\pi\right] (p^{+})^{\alpha-1} ( 1 - p^{+} )^{\alpha-1} \over (p^{+})^{2 \alpha} + 2 ( p^{+} )^\alpha ( 1 - p^{+} )^\alpha \cos(\pi \alpha) + (1 - p^{+} )^{2 \alpha}} .
\label{eq01}
\end{equation} 
When $\alpha=1/2$ the mentioned arcsine PDF is obtained which has
a $\cup$ shape. 
When $\alpha\to 1$
the PDF Eq. (\ref{eq01}) 
approaches a delta function on its mean $\langle p^{+} \rangle=1/2$
which corresponds to ergodic behavior. 
In the opposite limit $\alpha \to 0$ we have two delta functions on
$p^{+} = 0$  and $p^{+}=1$. Then the QD is  
either in the dark state or the bright state
for the whole duration of the measurement.
An important feature of the Lamperti PDF is that for $0<\alpha<1$ the maximum 
of the PDF is on these rare event: the PDF diverges on $p^{+}=0$ and
$p^{+}=1$. This clearly reflects sojourn times in either the on
or the off state which are of the order of the measurement time.

\begin{figure}
\begin{center}
\epsfig{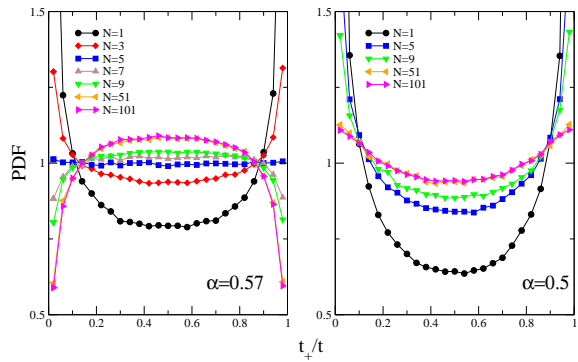}
\end{center}
\caption{ 
 Left panel $\alpha=0.57$ the PDF of residence fraction exhibits 
a  transition
between a $\cup$ to $\cap$ shape PDF which is controlled by  
$N$.  
In contrast such a transition is not observed for $\alpha=0.5$
and for smaller values of $\alpha$ (not shown).  
}
\label{fig2}
\end{figure}

{\em The PDF of the residence fraction} was obtained numerically
and is presented in Fig. \ref{fig2}
for $\alpha=0.57$ and $\alpha=0.5$. For
$\alpha=0.57$ a transition from a $\cup$ shape PDF for $N<5$ to 
a $\cap$ shape PDF for $N>5$ is found, while $N=5$ gives a uniform
PDF. 
 In contrast, for $\alpha=0.5$ the PDF for all $N$
has a $\cup$ shape. These findings indicate: (i) even in the
large $N$ limit the fluctuations of residence times remain non-trivial,
namely we do not see a convergence of the PDF to a delta function and (ii)
 that 
there exists a critical $\alpha$ below which the PDF of residence fractions
 has always (for any $N$)
a $\cup$ shape.
 However, simulations are limited to finite $N$ and finite measurement
time, hence we now turn to analytical theory.

 {\em The residence fraction for $N$ QDs} is
\begin{equation}
p^{+} = {1 \over t} \int_0 ^t \theta\left[ {\cal I}_N (t') - N/2\right] {\rm d} t' 
\label{eq02}
\end{equation} 
where $\theta(x)$ is the step function: $\theta(x) = 1$ if its argument 
is positive, otherwise it is zero. For convenience we will consider 
only odd $N$. The states of the system are determined by a vector
$\left[ I_1(t), ... I_i(t) ... I_N(t)\right]$. States contributing to
the considered residence fraction are those with at least $(N+1)/2$
processes in bright state $I_i(t) = 1$ and we have $2^{N}/2$ such states. 
We can rewrite the residence
fraction as a sum over time averages of these states
\begin{equation}
p^{+} = {1 \over t} \int_0 ^t \sum_{j=0}^{{N-1 \over 2} }
  \Pi_{i=1} ^N I_i(t') \bullet ^j {\rm d} t'.
\label{eq03}
\end{equation} 
Here $\Pi_{i=1} ^N I_i \bullet ^j$ is a sum over all permutation
of the product $I_1...I_N$ with $j$ terms of the type $1- I_i$ and
 and $N-j$ terms of the type $I_i$.
For example for $N=3$ we have 
  $\Pi_{i=1} ^N I_i \bullet ^0=I_1 I_2 I_3$
for $j=0$ and
$\Pi_{i=1} ^N I_i \bullet ^1=(1-I_1) I_2 I_3+ I_1(1-I_2)I_3 + I_1 I_2 (1 - I_3)$.
This means that
$p^{+}$ has contributions from states $(1,1,1)$ (all dots bright)
or $(0,1,1),(1,0,1),(1,1,0)$ since for these  states the
total intensity ${\cal I}_{N=3}(t)$ is above its mean $\langle {\cal I}\rangle _{N=3} = 3/2$. 

\begin{table}[ht]
\centering
\begin{tabular}{ c c c}
\hline\hline
$N$ & $\sigma^2$ \\
\hline
\ $5$ & $36 C_5 -45 C_4 + 25 C_3 - { 15 \over 2} C_2 + {15 \over 8} C_1  - {1 \over 4}$ \\
\hline
\ $7$ & $400C_7 - 700 C_6 + 546 C_5 - 245 C_4 + 70 C_3 -{105 \over 8} C_2 + { 35 \over 16} C_1 - { 1 \over 4} $  \\
\hline
\ $9$ & $4900 C_9 - 11025 C_8 + 11250 C_7 - 6825 C_6 + { 10899 \over 4 } C_5 $\\ 
\ $\ $ & $ - {5985 \over 8} C_4 + {1155 \over 8} C_3 - {315 \over 16} C_2 + {315 \over 128} C_1 -{1 \over 4}  . $ \\
\hline
\end{tabular}
\caption{
Variance of the residence fraction $\sigma^2$
for various $N$. 
}
\label{table1}
\end{table}

{\em Sub and super uniform statistics.} It is easy to see that the mean is
$\langle p^{+} \rangle= 1/2$ which is expected from symmetry. More interesting
is the variance
$\sigma^2 = \langle \left( p^{+}\right)^2 \rangle - \langle p^{+} \rangle^2$.
The variance can be used to quantify 
different types of fluctuations. If $\sigma^2 > 1/12$ we define the
fluctuations as  super-uniform  while $\sigma^2<1/12$ is sub-uniform.
Distributions of residence fractions with $\cup$ $(\cap)$ shape are  
super-uniform (sub-uniform) respectively. 
Squaring Eq. (\ref{eq03}) and averaging with respect to the random process
 $$ \langle (p^{+} )^2 \rangle = $$
\begin{equation}
{1 \over t^2}  \sum_{l=0} ^{N-1} A_N(l)   \int_0 ^t {\rm d} t_1  \int_0 ^t {\rm d} t_2  \langle I(t_1) \left[ 1 - I(t_2)\right]\rangle^l \langle I(t_1) I(t_2) \rangle^{N-l}.   
\label{eq04} 
\end{equation}
We see that the fluctuation of the residence fraction
is determined by the intensity correlation function
$\langle I(t_1) I(t_2) \rangle$ (here  
$\langle I(t_1)\left[1-I(t_2)\right]\rangle=1/2-\langle I(t_1) I(t_2) \rangle$).
A lengthy combinatorial calculation, which we will publish elsewhere,
yields the coefficients $A_N(l)$
\begin{equation}
A_N(l) = \left\{ 
\begin{array}{c c} 
\binom{N}{l} 2^{N-1} \left[ 1 - S_N\left(l\right) \right] \ & \ \mbox{if} \ l \le { N - 1  \over 2} \\
\ & \ \\
\binom{N}{l}2^{N-1} S_N\left(N-l\right) \ & \ \mbox{if} \  l \ge { N-1 \over 2},
\end{array} 
\right.
\label{eq05} 
\end{equation}
where \cite{remark0}
%
\begin{equation}
  S_N(l) =   
2 \sum_{i=0} ^{\lfloor { l-1 \over 2} \rfloor} \binom{l}{i} {1 \over 2^{i}} {1 \over 2^{l-i}}  \sum_{\tilde{i}= {N+1\over 2} - l +i } ^{{ N-1 \over 2} - i }  \binom{N - l}{\tilde{i}} {1 \over 2^{N - l-\tilde{i}}} {1 \over 2^{\tilde{i}}}.
\label{eq12}
\end{equation} 
This representation will soon be
useful since we identify binomial expansions in it.

{\em Correlation functions.}
Many natural processes exhibit stationary behavior namely 
$\langle I(t_1) I(t_2) \rangle$ is a function of the time difference
$|t_2 - t_1|$.  However many other systems exhibit aging 
\begin{equation} 
\langle I(t_1) I(t_2) \rangle = {\cal C}(t_1/t_2) \ \ \ t_1/t_2<1,
\label{eq06} 
\end{equation} 
which is common for a system with infinite mean waiting time.
 For the two state QD process
\cite{Godreche,GenadPRL}
\begin{equation}
{\cal C}(z) = { 1 \over 4} \left[ { \sin \pi \alpha \over \pi} B(z;\alpha,1-\alpha) + 1\right] \ \ \ 0<z<1 
\label{eq07} 
\end{equation} 
and $B(z;\alpha,1-\alpha)= \int_0 ^z x^{\alpha -1} (1 - x)^{-\alpha} {\rm d} x$ is the tabulated
incomplete Beta function. 

\begin{figure}
\begin{center}
\epsfig{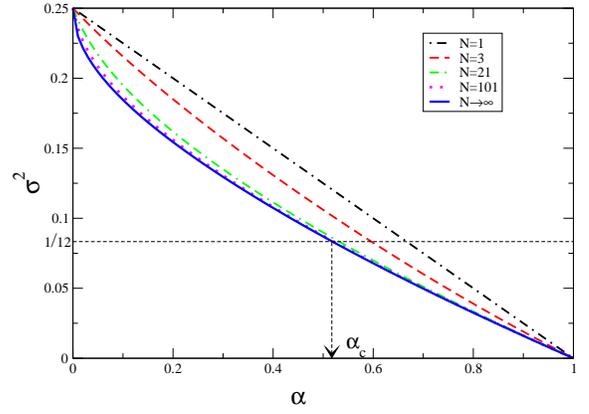}
\end{center}
\caption{ 
The variance of the residence fraction versus $\alpha$ for various number
of processes $N$. When $\alpha\to 1$ the process turns ergodic and
the variance is zero. For $\alpha<1$ even when $N \to \infty$ the 
fluctuations remain finite, indicating that in principle the non-ergodic
kinetics so far detected on the single particle level
can be found already for a large ensemble of particles.
Even
more surprisingly, for $N \to \infty$
 the fluctuations remain super-uniform, i.e. the variance is larger than $1/12$, when $\alpha<\alpha_c=0.518... $.
This implies that a $\cap$ shape PDF of the residence times  does not
describe fluctuations when $\alpha<\alpha_c$. 
}
\label{fig3}
\end{figure}

{\em Fluctuations for finite $N$.}  Inserting Eq. 
(\ref{eq07}) in Eq. (\ref{eq04}) changing variables to $z=t_1/t_2$ 
 we find exact expressions for the fluctuations.
For $N=1$ $\sigma^2 = 1-\alpha$, while 
 for 
$N=3$ 
\begin{equation} 
 \sigma^2 =  4 C_3 - 3 C_2 + {3 \over 2} C_1 -{1\over 4} 
\label{eq07a} 
\end{equation} 
where
\begin{equation}
C_n = \int_0 ^1 {\cal C}^n (z) {\rm d} z.
\label{eq08} 
\end{equation} 
Cases  $N=5,7,9$ are reported in Table 1 and 
similarly we obtain $\sigma^2$ for
any finite $N$ using the exact expression for $A_N(l)$. 
With Table 1 we can determine whether fluctuations are sub or super uniform.
For example for $N=5$ we find uniform statistics
 $\sigma^2=1/12$  when $\alpha\simeq 0.57$.
This  perfectly matches the simulations in Fig. \ref{fig2} which exhibit a
uniform distribution of the occupation fraction for these parameters.   
For $\alpha=1/2$, i.e.  Brownian motion,
Eqs. (\ref{eq07},\ref{eq08}) give
\begin{equation}
 C_n =  
\sum_{m=0} ^n \binom{n}{m} { (i \pi)^{-m} \over 4^{n+1}}  \left[ m! -\Gamma\left(1 + m, - i \pi\right) - \Gamma\left( 1 + m, i \pi\right) \right]. 
\label{eq09}
\end{equation} 
For $\alpha\neq 1/2$ we solve the integral Eq. (\ref{eq08}) with Mathematica 
\cite{remark1}
and with Table 1 obtain numerically exact expressions for $\sigma^2$.
In Fig. \ref{fig3} we show $\sigma^2$ versus $\alpha$ for various $N$.
When $\alpha=1$ we have
$C_n= \langle I \rangle^{2 n} = 4^{-n}$ and then using Table
1 we get $ \sigma^2 = 0$ which corresponds to the ergodic phase. In the
opposite limit $\alpha \to 0$ we have $C_n = \langle I \rangle^n = 2^{-n}$
and then $\sigma^2 = 1/4$. This is  the expected behavior since
when $\alpha \to
0$ a single particle remains in one state (either bright or dark)
for the whole measurement time and hence with probability $1/2$ 
the signal composed of an ensemble of an odd number
of  particles  is above its 
mean all along the measurement. 
As shown in Fig. \ref{fig3} 
the fluctuations are the strongest when $N=1$ which is expected.
Surprisingly, we see that as the number of particles is
increased the fluctuations remain finite. This means that
the non ergodic behavior found for a single QD \cite{Dahan,GenadPRL}
 can be detected
also for a large ensemble of dots. A measurement of the residence
time $t^{+}$ 
remains a random variable even when $N$ is large and
a single measurement is not a  reproducible result.
 We now turn to find the variance
$\sigma^2$ in the $N\to \infty$ limit.

{\em Limit $N\to \infty$.}
The limit of large $N$ and $l$ is taken in such a way that $x=l/N$
 remain finite.
 We then use Stirling's formula to approximate
the double binomial  sum in Eq. (\ref{eq12})
with a double integral over a pair of Gaussians. The first 
integration yields an error function and we find
\begin{equation}
S_N(l) \sim 2 \int_0 ^\infty \sqrt{ x N \over 2 \pi} e^{ - { N x z^2 \over 2} } \mbox{Erf}\left( { \sqrt{N} x z \over \sqrt{2} \sqrt{ 1 - x}} \right) {\rm d} z .  
\label{eq13}
\end{equation} 
From a table of integrals  
$S_N (l)  \sim { 2 \over \pi} \mbox{ArcSin}\left( \sqrt{x} \right)$ and hence we
find
\begin{equation}
A_N (l) \sim \binom{N}{l} 2^{N-1} \left[ 1 - {2 \over 
\pi} \mbox{ArcSin}\left( \sqrt{ {l \over N}} \right) \right]. 
\label{eq15} 
\end{equation}
Inserting this expression in
 Eq. (\ref{eq04})  and using 
Eq. (\ref{eq06}) we find (after change of variables to $z=t_1/t_2$)
$$ \langle (p^{+} )^2 \rangle \sim $$ 
\begin{equation}
\int_0 ^1 {\rm d} z \sum_{l=0} ^{N} {1 \over 2} \left[ 1 - {2 \over \pi}
\mbox{ArcSin} \sqrt{ {l \over N} } \right] \binom{N}{l} \left[
1 - 2 {\cal C}(z) \right]^l \left[ 2 {\cal C}(z) \right]^{N-l} . 
\label{eq16} 
\end{equation}  
In the large $N$ limit the binomial part of the integral:
$
\binom{N}{l} \left[
1 - 2 {\cal C}(z) \right]^l \left[ 2 {\cal C}(z) \right]^{N-l}$, is narrowly
centered around its
mean $\left[ 1 - 2 {\cal C}(z) \right] N$. Hence  
we find
\begin{equation}
\lim_{N \to \infty} \sigma^2  = {1 \over \pi} \int_0 ^1 {\rm d} z \mbox{ArcSin} \sqrt{ 2 {\cal C}(z) } - {1 \over 4},
\label{eq17} 
\end{equation} 
which is the main equation of this manuscript.

Eq. (\ref{eq17}) shows that  even  when $N\to \infty$ the statistics
of blinking QDs is non
ergodic: the fluctuation of the residence time are 
finite even for a large ensemble. 
Only in the limit $\alpha \to 1$ we do get expected ergodic behavior
$\sigma^2 = 0$. 
  In Fig. \ref{fig3} we plot Eq. (\ref{eq17}) versus $\alpha$
and classify sub-uniform and super-uniform statistics. 
 For the Brownian
case $\alpha=1/2$ we get $\lim_{N \to \infty} \sigma^2 = 0.086\cdots$
which is super-uniform, hence this rigorously shows that 
a $\cap$ shape PDF is not
found \cite{remark2}. 
L\'evy's arcsine law exhibits super-uniform statistics
and surprisingly this property is maintained 
also for a large ensemble of particles. 
 Inserting Eq. (\ref{eq07}) in Eq. (\ref{eq17}) we
find that if $\alpha< \alpha_c= 0.518\cdots$ 
the fluctuations are super-uniform in the limit $N \to \infty$ 
(see Fig. \ref{fig3}).
This theoretical prediction is in agreement our simulations
presented in Fig. \ref{fig2} which
shows super-uniform (sub-uniform) behaviors when $\alpha<\alpha_c$
($\alpha> \alpha_c$) respectively. 


 Bouchaud \cite{B92},
introduced the profound concept
of weak ergodicity breaking 
 in the context of glassy dynamics. 
It holds for
systems
with power law sojourn times 
and with a phase space  not  broken into mutually
inaccessible regions, e.g. blinking QDs.
For a macroscopic system like a glass
Bouchuad assumed that  measurements
are made on a large number of sub-systems (micro spin-glasses of clusters of magnetic atoms in his case)
and that such measurements are disorder averaged.  He claimed
that performing dynamical experiments on {\em  mesoscopic samples, where $N$
is small}, would yield irreproducible results. 
Indeed we thoroughly believed this scenario at the start of this project,
namely our  naive expectation was that as the number of
 QDs increases we
will approach an ergodic behavior. 
However, it turns out that a measurement of
the residence time of $N \to \infty$
QDs is irreproducible. Weak ergodicity breaking can be detected
for a large ensemble of sub-systems. 

{\bf Acknowledgment} This work was supported by the Israel science foundation.


\begin{thebibliography}{99}
\bibitem{remark1} Properties of $C_n$ will be investigated elsewhere,
e.g. for large $n$ steepest descent gives
 $ C_n \sim 
  \Gamma\left({2-\alpha  \over 1 - \alpha} \right) \left[ {  \sin \pi \alpha \over \left( 1 - \alpha \right) 2 \pi } \right]^{-{1 \over 1 - \alpha}} \left( {1 \over 2} \right)^n n^{ - {1 \over 1 - \alpha}} $.
\bibitem{remark2} Our exact expression for 
$\sigma^2$ cannot be used to determine the existence of 
singularities in the PDF of $t^{+}/t$. From simulations we find
that $\alpha\simeq 1/2$ is a critical point in the sense that
these singularities vanish in the $N \to \infty$ limit.
\end{thebibliography}
\end{document}